# RESONANT X-RAY SCATTERING FROM THE SURFACE OF A DILUTE LIQUID Hg–Au ALLOY

E. DiMasi,[1*] H. Tostmann,[2†] B. M. Ocko,[1] P. Huber,[2] O. G. Shpyrko,[2] P. S. Pershan[2], M. Deutsch[3], and L. E. Berman[4]

[1]Department of Physics, Brookhaven National Laboratory, Upton NY 11973
[2]Div. of Eng. and Appl. Sci. and Dept. of Physics, Harvard University, Cambridge MA 02138
[3]Department of Physics, Bar-Ilan University, Ramat-Gan 52100, Israel
[4]National Synchrotron Light Source, Brookhaven National Laboratory, Upton NY 11973
*Corresponding author: dimasi@bnl.gov
†Present address: Department of Chemistry, University of Florida, Gainesville FL 32611

## ABSTRACT

We present the first resonant x-ray reflectivity measurements from a liquid surface. The surface structure of the liquid Hg-Au alloy system just beyond the solubility limit of 0.14at% Au in Hg had previously been shown to exhibit a unique surface phase characterized by a low-density surface region with a complicated temperature dependence. In this paper we present reflectivity measurements near the Au $L_{III}$ edge, for 0.2at% Au in Hg at room temperature. The data are consistent with a concentration of Au in the surface region that can be no larger than about 30at%. These results rule out previous suggestions that pure Au layers segregate at the alloy surface.

## INTRODUCTION

Surfaces play an important role in nucleating bulk phases in solid alloys, and because of the reduced atomic coordination, surfaces often exhibit structure and phase behavior distinct from that of the bulk. Liquid metal alloys can be expected to exhibit an even richer variety of surface phases: since the particles are not constrained to lattice sites, the composition at the surface can vary dramatically from that of the bulk. With the application of surface sensitive x-ray scattering techniques to the liquid metal–vapor interface, a number of liquid metal alloys have in fact been shown to exhibit interesting structural phenomena, such as surface segregation and microscopic precursors to wetting, that previously could be only theoretically predicted or indirectly inferred from experiment.[1,2]

Especially interesting behavior was found for liquid Hg-Au alloys near the room temperature solubility limit of 0.14at% Au in Hg.[3] The complicated phase behavior of Hg-Au alloys has long been known, from bulk structural studies,[4] and from studies at the solid amalgam surface.[5] Further motivation for studying liquid Hg alloys comes from experiments in which metal impurities were observed to affect the activation energies of reactions catalyzed by the liquid Hg surface.[6] It is not known whether such effects are due to changes in the electronic properties of liquid Hg, modification of the surface structure, or the formation of intermetallic phases at the surface.

Our study of the x-ray reflectivity of the Hg-Au system at temperatures between $-39°$C and $+25°$C revealed two distinct regions of surface phase behavior, depending on whether the bulk solubility limit[7] of Au in Hg is exceeded. At high $T$ and low Au concentrations, surface layering similar to that of pure Hg is observed (Fig. 1, ○).[8] By contrast, at low $T$ and comparatively higher Au concentrations, reflectivity measurements revealed a more complicated surface phase, consistent with a low-density layer appearing at the interface (Fig. 1, •).

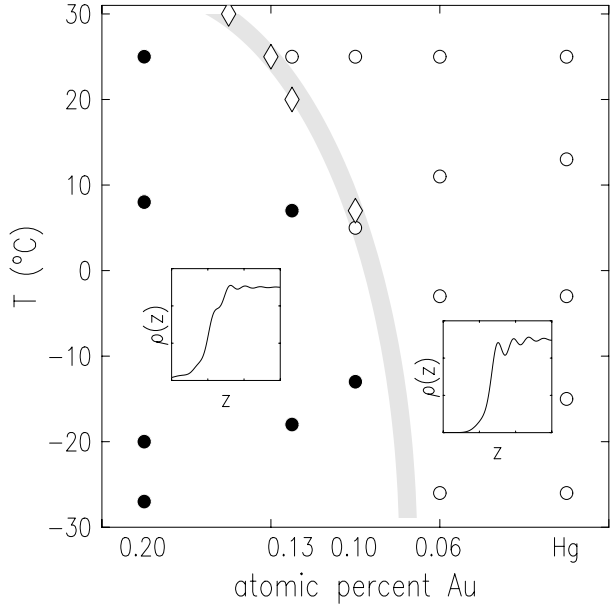

Figure 1: Hg-Au surface phase diagram established in Ref. 3. ◇: Measured solubility limit from Ref. 7 (shaded band is a guide for the eye). ○: Compositions and temperatures for which reflectivity indicates simple oscillatory real-space density profiles, as the inset on the right. •: Compositions and temperatures exhibiting less well defined surface layering and a low-density surface region (inset, left).

Since the new phase was found only for Au-rich samples, it is natural to ask how the Au composition in the surface region differs from that of the bulk. Unfortunately, this elemental specificity is difficult to obtain experimentally. X-ray reflectivity, in which scattered x-ray intensity is measured as a function of momentum transfer $q_z$ normal to the surface, is a sensitive probe of the surface-normal electron density distribution $\rho(z)$, and in the kinematic limit may be taken to be proportional to the Fresnel reflectivity $R_F$ of a homogeneous surface:[9]

$$R(q_z) = R_F \left| (1/\rho_\infty) \int_{-\infty}^{\infty} (\partial \rho / \partial z) \exp(i q_z z)\, dz \right|^2, \qquad (1)$$

where $\rho_\infty$ is the density of the bulk. The electron density variations that appear as modulations in the reflectivity may result either from a compositional change or a mass density change. Thus, direct determination of composition from reflectivity is ambiguous at best. Complementary information is often obtained from electron spectroscopy techniques requiring ultra high vacuum conditions, which cannot be used with high vapor pressure liquid Hg alloys.

These disadvantages can be overcome with the application of resonant x-ray scattering. The apparent electron density of a scattering atom depends on the scattering form factor, which can be taken as $f(q) + f'(E) \approx Z + f'(E)$, where absorption and the weak $q$ dependence have been neglected. When the x-ray energy is tuned to an absorption edge of a scattering atom, the magnitude of $f'$ becomes appreciable. The resulting change in the contrast between unlike atoms provides composition information for the model structure.[10]

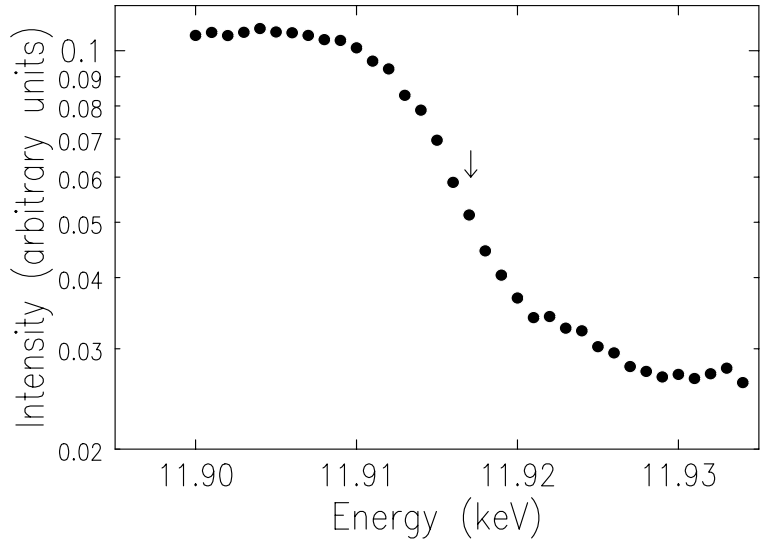

Figure 2: Transmission of x-rays through an Au foil as a function of energy near the Au $L_{III}$ edge (●). The arrow indicates the energy selected for reflectivity measurements.

The starting point for the analysis we will use for this report is a simple real-space density model of the region within 5 Å of the surface:

$$\frac{\rho(z)}{\rho_\infty} = \frac{1}{2}\mathrm{erf}\left(\frac{z}{\sigma_0\sqrt{2}}\right) + \frac{\rho_1/\rho_\infty}{2}\left[\mathrm{erf}\left(\frac{z-z_1}{\sigma_1\sqrt{2}}\right) - \mathrm{erf}\left(\frac{z}{\sigma_0\sqrt{2}}\right)\right] + \frac{h_g/\sigma_g}{\sqrt{2\pi}}\exp\left[\frac{-(z-z_g)^2}{2\sigma_g^2}\right]. \quad (2)$$

The three terms define, respectively, the liquid–vapor interface at $z = 0$, a region of density $\rho_1$ at position $z_1$, and a broad density tail further towards the vapor region. Because of the very small bulk concentration of Au, the reflectivity is calculated as though the system is composed entirely of Hg. A more complete discussion of the modeling is given in Ref. 3.

## EXPERIMENT

Au powder was dissolved in pure liquid Hg for a nominal composition of 0.2at% Au. The samples were contained in an ultra high vacuum chamber, with the pressure determined by the Hg vapor pressure at room temperature, with low partial pressures of oxygen and water. The experiment was performed at beamline X25 at the National Synchrotron Light Source using a specialized BNL-designed liquid surface spectrometer. Details of the sample preparation and measurement technique have been given previously,[3] except that for this experiment, a vertically deflecting Si(111) double crystal monochromator upstream of the spectrometer was used to improve the energy resolution to $\sim 10$ eV.

For the present work, we compare reflectivity measured at 11 keV to measurements made at the Au $L_{III}$ edge at 11.919 keV. The energy choice was based on transmission of the direct beam through an Au foil, shown in Fig. 2.[11] At the inflection point indicated by the arrow, $f'_{\mathrm{Au}}$ has its minimum value of $-19$ electrons, compared to about $-6$ electrons in the 10–11 keV region.[10] Here the scattering from Au (with $Z=79$) is reduced to about 85% of its 11 keV value. Because of the very small bulk concentration of Au, no Au fluorescence was detected at any energy employed.

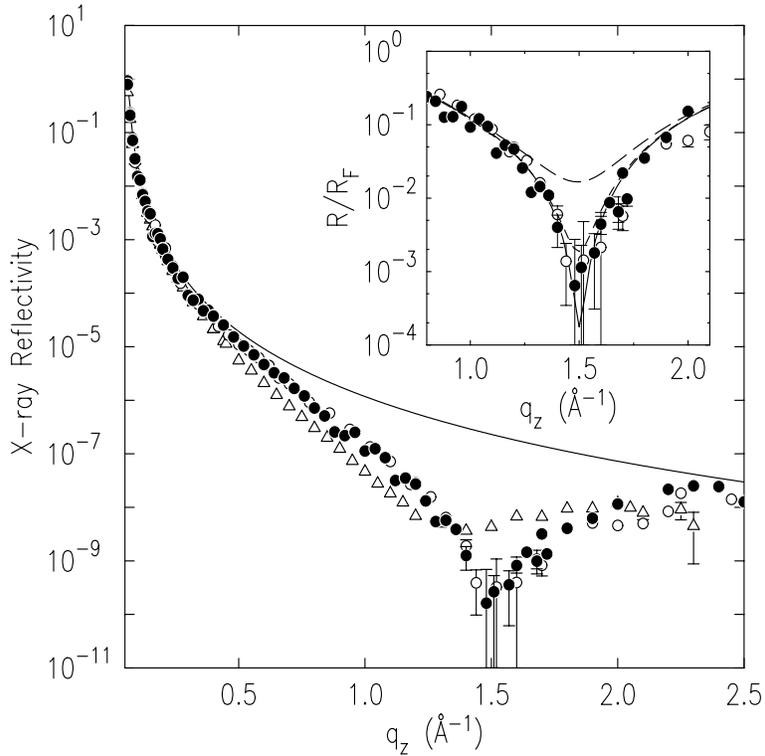

Figure 3: X-ray reflectivity vs. $q_z$ for Hg 0.2at% Au reported in Ref. 3 at 10 keV ($\triangle$), and for the present sample at 11 keV ($\circ$) and 11.919 keV ($\bullet$). Solid line: calculated Fresnel reflectivity $R_F$ of a flat Hg surface. Inset: reflectivity data normalized to $R_F$ at 11 keV ($\circ$) and 11.92 keV ($\bullet$). Lines are calculated from models as described in the text.

**RESULTS**

X-ray reflectivity obtained from the Hg 0.2at% Au sample at 11 keV (Fig. 3, $\circ$) is similar to that obtained on a sample of approximately the same composition measured previously at 10 keV (Fig. 3, $\triangle$). Compared to the Fresnel reflectivity $R_F$ calculated for a flat Hg surface (Fig. 3, solid line), the experimental data have deep minima for $q_z$ in the range 1.3–1.5 Å$^{-1}$. This destructive interference comes from a region near the surface where the density is approximately half that of the bulk liquid. Since the previous study showed that the details of the surface structure are very sensitive to Au concentration in this region, the differences between the two samples can be ascribed to a slight difference in the Au concentration.

The normalized reflectivity data have been described by the model profile shown in Fig. 4 (solid line), which produces the calculated reflectivity shown as a solid line in the inset of Fig. 3. The local minimum in the density between the surface layer and the bulk, at $z = -0.5$ Å, is due to the rather small roughnesses used in this simple model, which are required in order to produce the relatively large reflectivity we observe for $q_z > 1.8$ Å$^{-1}$.

Essentially no difference is seen in the data taken at the Au $L_{III}$ edge. This rules out models having a very high Au composition at the surface. To illustrate the sensitivity of the measurement to the Au composition in the density profiles, we have calculated reflectivity curves for two additional cases. If the entire surface region ($z < 0$) were composed of Au, the electron density contributing to the scattering would be reduced by 15% at 11.919 keV. This density profile (Fig. 4, long dashed line) produces a reflectivity curve with a much less pronounced minimum at 1.5 Å$^{-1}$ Fig. 3 (inset, long dashed line). A similar calculation assuming 40% Au in the surface layers, indicated by short dashed lines, lies just outside the experimental error.

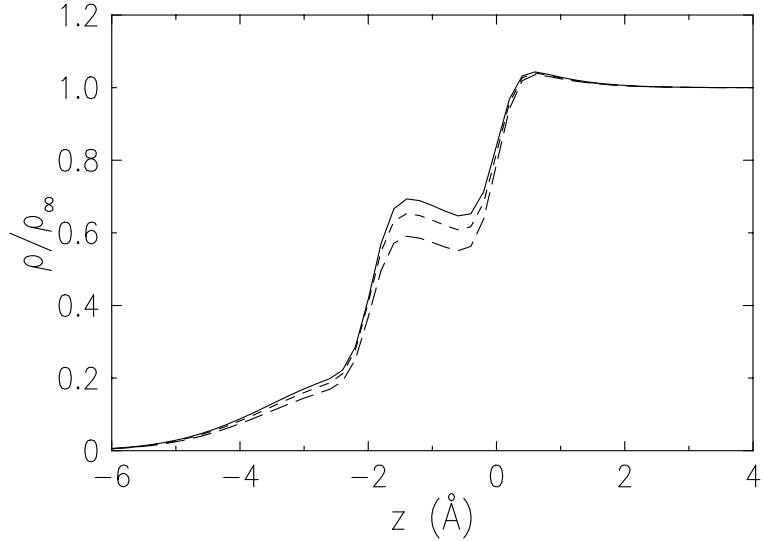

Figure 4: Surface density profiles calculated from the model in Eq. 2. (—): "Pure Hg" profile: $\sigma_0 = \sigma_1 = 0.25$, $z_1 = -1.96$, $\rho_1 = 0.51$, $h_g = 0.8$, $\sigma_g = 1.5$, and $z_g = -2$. (- - -): "40% Au" surface layer, where the apparent density of the step region is reduced by 6% ($\rho_1 = 0.47$, $h_g = 0.75$). (– – –): "100% Au" surface layer, where the apparent density of the step region is reduced by 15% ($\rho_1 = 0.43$, $h_g = 0.68$).

A surface composition of 30% Au is compatible with our data (model not shown), and forms an upper bound on the Au concentration, assuming that the energy was tuned precisely to the minimum in $f'$. To reduce the change in $f'$ by a factor of two, yielding 60% as the upper bound for Au concentration, an error in the energy of $> 30$ eV would be required.[10] This error is large compared to the width of the energy scan shown in Fig. 2. Further confidence in our experimental sensitivity comes from a subsequent experiment we performed on a liquid Bi-In alloy at the Bi $L_{III}$ edge, where large resonant effects were observed.[12] Our results rule out models in which essentially pure Au layers segregate at the surface of the alloy and produce a coherent contribution to the reflectivity.

## CONCLUSIONS

We present the first resonant x-ray scattering measurements from a liquid surface. The surface structure of the liquid Hg-Au alloy system just beyond the solubility limit of 0.14at% Au in Hg had previously been shown to exhibit a unique Au-rich surface phase characterized by a low-density surface region with a complicated temperature dependence. In this paper we present reflectivity measurements near the Au $L_{III}$ edge. The data are consistent with a concentration of Au in the surface region that can be no larger than about 30at%. Our results rule out models in which the surface region consists of pure Au. Since a concentration of ≤ 30at% Au at the surface is consistent with the data, the surface phase may still be quite different from the 0.2at% composition of the bulk.

## ACKNOWLEDGMENTS

We are indebted to Scott Coburn of BNL Physics for the design, construction, and preparation of the liquid surface spectrometer that made these experiments possible. We acknowledge support from the U.S. DOE (DE-FG02-88-ER45379, DE-AC02-98CH10886), the U.S.–Israel Binational Science Foundation, and the Deutsche Forschungsgemeinschaft.